\documentclass[aps,prl,twocolumn,showpacs, superscriptaddress]{revtex4-1}

\usepackage{graphics}
\usepackage{amsmath}
\usepackage{wrapfig}
\usepackage{epsfig}
\usepackage[utf8x]{inputenc}
\usepackage{wasysym}
\usepackage{verbatim}    
\usepackage{hyperref}
\usepackage{psfrag}
\usepackage{color}
\usepackage{natbib}
\usepackage{bm}

\renewcommand{\vec}[1]{\mathbf{#1}}

\begin{document}

\title{Nucleation-induced transition to collective motion in active systems}

\author{Christoph A. Weber}
\affiliation{Arnold Sommerfeld Center for Theoretical Physics and Center for NanoScience, Department of Physics, Ludwig-Maximilians-Universit\"at M\"unchen, 80333 Munich, Germany}
\author{Volker Schaller}
\affiliation{Lehrstuhl f\"ur Biophysik (E27), Technische Universit\"at M\"unchen, 84748 Garching, Germany}
\author{Andreas R. Bausch}
\affiliation{Lehrstuhl f\"ur Biophysik (E27), Technische Universit\"at M\"unchen, 84748 Garching, Germany}
\author{Erwin Frey}
\affiliation{Arnold Sommerfeld Center for Theoretical Physics and Center for NanoScience, Department of Physics, Ludwig-Maximilians-Universit\"at M\"unchen, 80333 Munich, Germany}

%\date{\today}

\begin{abstract}
While the existence of polar ordered states in active systems is well established, the dynamics of the self-assembly processes are still elusive. We study a lattice gas model of self-propelled elongated particles interacting through excluded volume and alignment interactions, which shows a phase transition from an isotropic to a polar ordered state. By analyzing the ordering process we find that the transition is driven by the formation of a critical nucleation cluster and a subsequent coarsening process.  Moreover, the time to establish a polar ordered state shows a power-law divergence.
\end{abstract}

\maketitle

Understanding collective motion in driven or self-propelled particle systems is a topic of recent interdisciplinary interest \cite{Vicsek_Review, Ramaswamy_Review, Aranson_Tsimring_book}. Coherently moving groups have been observed over a broad range of scales, spanning from micrometer-sized systems~\cite{Butt, Schaller, schaller2, Dombrowski_2004, Ringe_PNAS, Yutaka, pnas_bacteria}   over millimeter-large granules~\cite{Dachot_Chate_2010, Dauchot_long, Kudrolli_2008}  to large cooperative animal groups~\cite{ballerini_starlings}. The ubiquity of this phenomenon raises the questions of how these coherently moving and ordered clusters arise. Coarse-grained particle-based models~\cite{Vicsek, Bussemaker, Chate_2004, Chate_Variations, Chate_long, Albano2009,  Peruani_Traffic_Jams, Ginelli} have shown that interactions favoring the \emph{alignment} of the particles' direction of motion is sufficient for the emergence of large scale order. As possible origins for the alignment, excluded volume interactions~\cite{Peruani_rods}, dissipative collisions~\cite{Grossman_Aranson} and hydrodynamic interactions \cite{Graham_2005, Graham_2009} have been discussed.Complementary to these coarse-grained agent-based models, hydrodynamic models have been derived from mesoscopic collision rules \cite{Aronson_MT,Bertin_short,Aranson_bacteria_2007,Bertin_long,Ihle_2011}, or by means of microscopic interactions \cite{Saintillan_2007, Saintillan_2008, Marchetti2008,Baskaran_Marchetti_2008,Mishra_Marchetti_2010}.
  These approaches allow to determine the kinetic coefficients, and thereby to analyze pattern forming instabilities. Furthermore, hydrodynamic equations based on symmetry arguments \cite{Toner_Tu_1995, Toner_Tu_1998, Ulm_Toner_Tu_1998, Ramaswamy_swimmers_2002, GNF_substrate, RamaswamyReview} were derived, enabling to show that collections of self-propelled particles exhibit a true, long-range ordered, spontaneously broken symmetry state. In addition, scaling exponents have been calculated analytically \cite{Toner_Tu_1995, Toner_Tu_1998},  and validated by agent-based simulations \cite{Ulm_Toner_Tu_1998}.  Moreover, these studies revealed the existence of giant number fluctuations and long-lived density correlations, which have also been found experimentally \cite{GNF_in_granules,Dachot_Chate_2010,pnas_bacteria}.
\begin{figure}[!b]
\centering
\includegraphics[width=9cm]{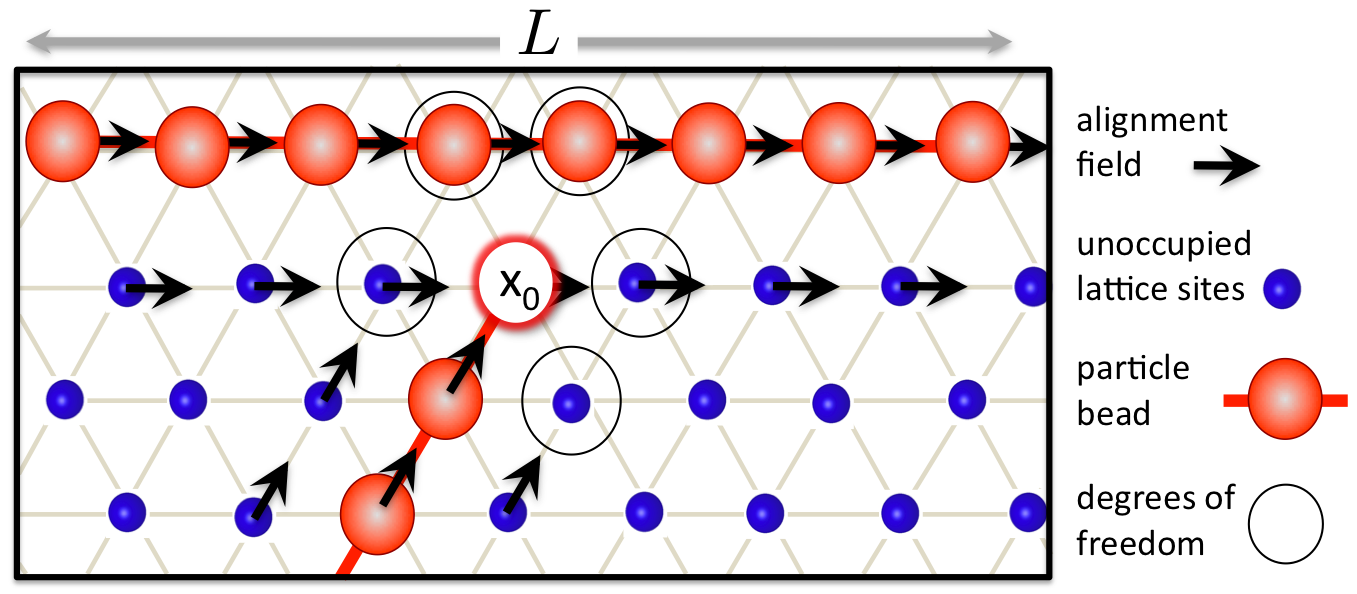}
\caption{\label{CA}
(color online) Illustration of the lattice gas model. Filaments of length $L$ move on a hexagonal lattice with constant speed performing a persistent random walk. %Each particles' direction performs a 
%persistent random walk with their heads ${\bf x}_0$ moving forward probabilistically to five different directions, indicated by black circles. %Excluded volume and alignment interactions are modeled by means of modified probabilities to move to preoccupied sites or to sites with a non-zero alignment field.
}
\end{figure}

While previous approaches on propelled particle systems focussed on the long-time dynamics, the assembly processes leading to collective motion remain elusive. To close this gap, we present an agent--based model of (self-)propelled particles interacting via an effective excluded volume interaction and a local polar alignment field. In the absence of excluded volume interactions, we find a phase transition to collective motion above a critical packing fraction. Strong excluded volume interactions increase this critical packing fraction or even inhibit the development of collective motion. Analyzing the dynamics of the pattern forming processes we find that collective motion is accompanied by a gain in free volume. Close to the critical packing fraction, the onset of collective motion requires the spontaneous formation of a cluster of sufficiently large mass that acts as a nucleus and triggers the transition to collective motion. The corresponding lag-time shows a power-law divergence upon approaching the critical packing fraction.

To follow the self-assembly processes over long time periods and for a large number of particles we consider a lattice gas automaton; cf. Fig.~\ref{CA}. 
Each of the $N$ particles is taken as a filament of length $L$ moving at constant speed $v$, and with a direction performing a persistent random walk. To emulate a spatially isotropic system as close as possible we choose a hexagonal lattice, and use a simulation box of hexagonal shape to avoid possible artifacts due to symmetry breaking with the local hexagonal lattice structure; all length are measured in units of the lattice constant. The time scale is fixed by allowing each filament's head to move one lattice unit per time step. The particles' dynamics are assumed to be fully determined by its head, while the tail strictly follows the head's trail. The persistent random walk is implemented by a stochastic process, where each filament's head ${\vec x}_0$ moves to one of its five neighboring sites ${\vec x}_\alpha$ according to a fixed set of conditional probabilities $P_\text{prw} ({\vec x}_\alpha | \vec x_0)$.  
For a fixed filament length $L=10$ we choose the conditional probabilities such that each filament performs a persistent random walk with constant unit speed and a kinetic persistence length $\ell_p^0=8.7$ comparable to $L$; for more details see the Supplementary Material.  

Our model includes an effective excluded volume interaction and a local alignment field. Both enter into the model by supplementing the probabilities for the unperturbed persistent random walk by appropriate factors. In contrast to recent off-lattice simulations of over-damped rods, solely interacting by means of excluded volume~ \cite{Peruani_rods}, this allows us to separately tune excluded volume and alignment interaction and thereby study their interplay. For a lattice site that is already $k$-times occupied,  the probability for further occupation is reduced by a Boltzmann factor, $e^{- k  \, \epsilon }$, where $\epsilon$ characterizes the penalty for multiple occupations. Formally, for $\epsilon \to  \infty$, the limit of strict excluded volume  is obtained \cite{Marchetti2008}. The limit of 
weak penalties for multiple occupations  (low $\epsilon$) is appropriate for motility assay experiment, where filament crossings occur  frequently \cite{Schaller, schaller2, Ringe_PNAS, Yutaka}.  Moreover, these experiments also indicate that there is a local alignment interaction between filaments~\cite{Schaller, schaller2, Ringe_PNAS, Yutaka}. To emulate such an interaction, each particle is assigned an alignment field $u(x,t)$ at its occupied and neighboring lattice sites (see arrows in Fig.1); it is directed along the particles contour, and is of unit length. Overlapping alignment fields of different particles are averaged. In a collision event 
%with another particle 
the alignment field modifies the transition probability to move from $\vec{x}_0$ to $\vec{x}$ by the Boltzmann factor, $\alpha^{\cos \varphi}$, where $\varphi$ is the relative angle between the alignment field and the direction of motion of the respective collision partner, $\varphi=\angle(\vec u(\vec x), \vec x - \vec  x_0)$.  The parameter $\alpha \ge 1$ characterizes the strength of the polar alignment, where stronger interactions correspond to increasing values of $\alpha$.  Assuming that all these contributions to the filament's dynamics are statistically independent, we arrive at the following update rule: Given a configuration of $N$ filaments we use random sequential updating and move a chosen head from position $\vec x_0$ to a target position $\vec x$ with probability
$P(\vec x | \vec x_0; \vec u, k) \propto P_{\text{prw}}(\vec x | \vec x_0) \cdot e^{- k \, \epsilon}  \cdot  \alpha^{\cos \varphi(\vec{u}) }
$.

We have performed extensive numerical simulations for a broad range of reduced densities $\rho = N L / A$, where $A$ is the area of the simulation hexagon. For low densities an initial statistically homogeneous distribution of filaments remains isotropic both in filament position and orientation. However, for densities above some threshold, an initially disordered state evolves into a state with small coherently moving clusters that spread perpendicular to their direction of motion and form bands (see Fig.~\ref{Fig2}a-c, and videos in Supplementary Material).  The clusters move coherently embedded in an isotropic background of randomly oriented particles. These observations resemble similar results found previously in Vicsek-like models~\cite{Chate_long} and other agent-based lattice gas models~\cite{Bussemaker, Peruani_Traffic_Jams}. 

\begin{figure}[t]
\centering
\includegraphics[width=9cm]{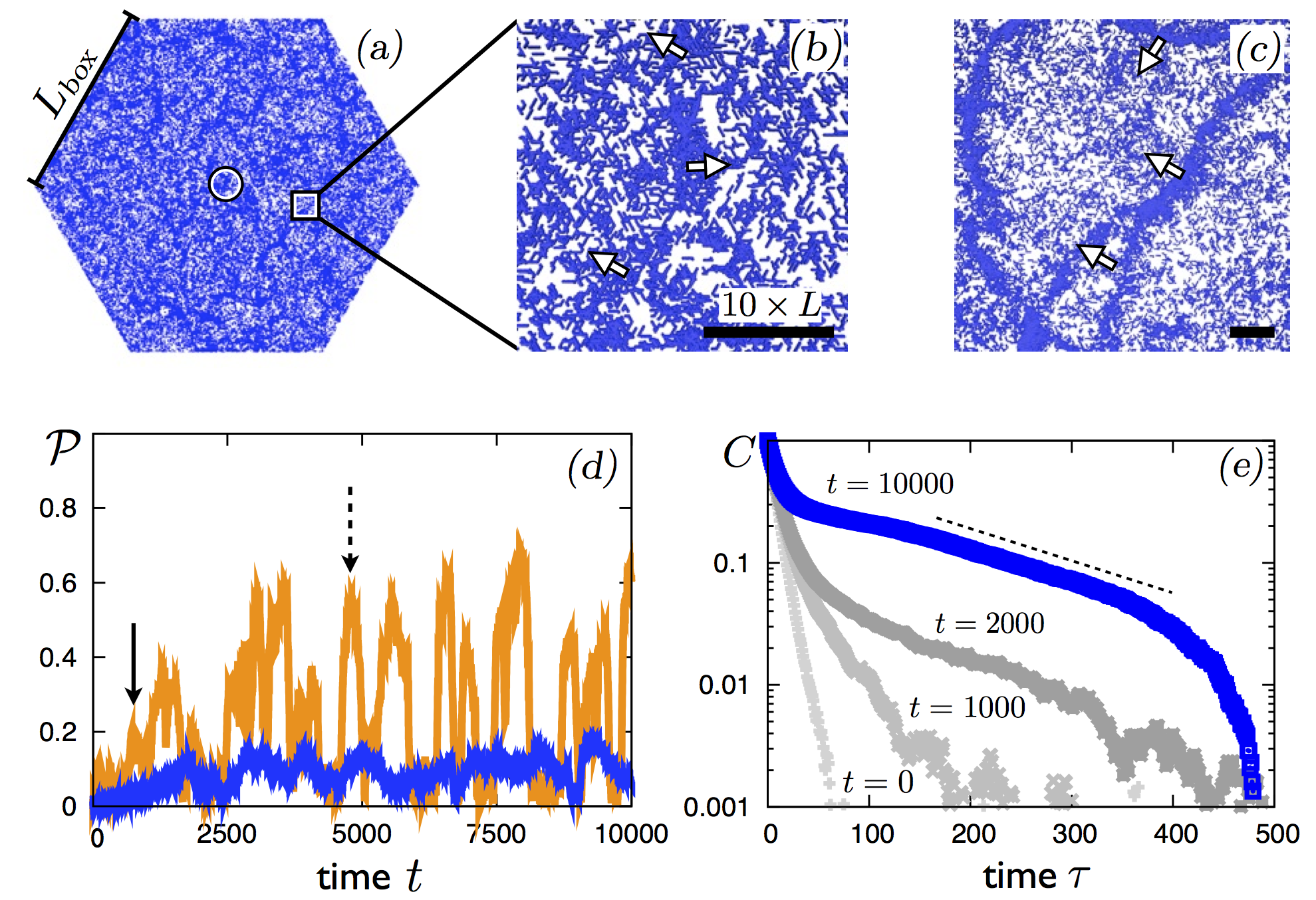}
\caption{\label{Fig2} (color online) {\bf (a)} The boundary for our lattice gas model is a reflecting hexagon with a side length $L_\text{box} = 1000$. The  circle indicates the ROI with a radius of $L_\text{box}/10$. Starting from a disordered state first small clusters emerge {\bf  (b)} which subsequently evolve into bands {\bf (c)} (see also videos in Supplementary Material). Directions of cluster movement are indicated by arrows. {\bf (d)}  The polarization ${\cal P}_\text{ROI}$ [orange, light grey]  is calculated over all particles in the ROI, while the polarization ${\cal P}_\text{SIM}$ [blue, dark grey] is determined over all particles in the entire simulation box. The bold and dashed arrows correspond to snapshots (b) and (c), respectively. {\bf (e)} Orientational correlation function $C(\tau; t)$ as a function of time $\tau$ for a series of initial times $t$ as indicated in the graph. Simulation results refer to the parameters $\alpha=5$, $\epsilon=0.5$, and $\rho=0.6$. All simulations were typically run for  $\sim 2 \cdot 10^{4}$ time steps.}
\end{figure}

\begin{figure*}[!t]
\centering
\includegraphics[width=18cm]{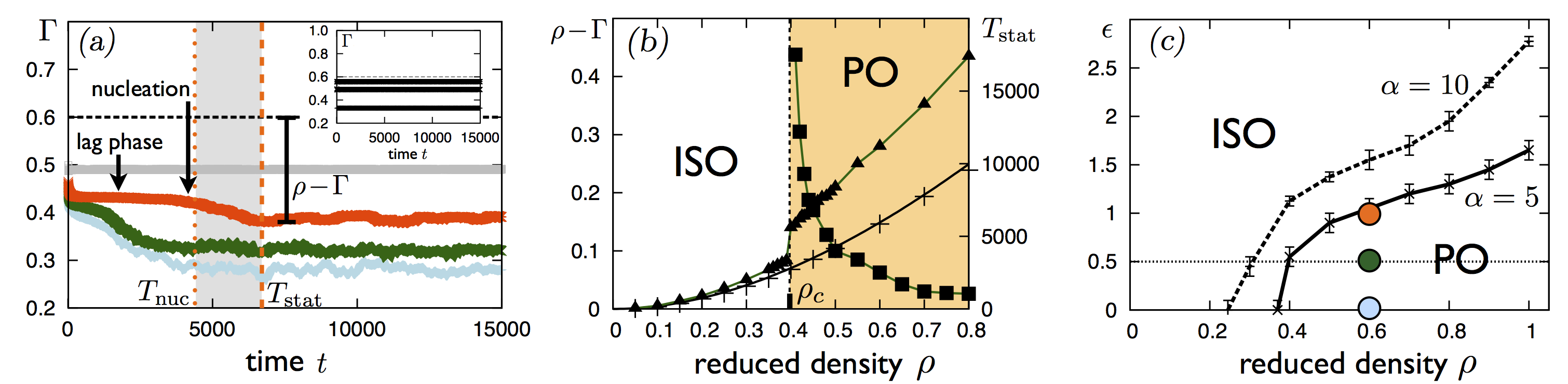}
\caption{\label{fig:transition}
(color online)
{\bf (a)} \emph{Time traces} of  $\Gamma$ for $\rho=0.6$, and a set of values: $\epsilon=0, 0.5, 1$ with $\alpha = 5$, and $\epsilon=1$ with $\alpha=1$, from bottom to top. Close to the phase boundary from the isotropic (ISO) to the polar ordered state (PO) [$\epsilon=1$; (red)], a lag-phase exists, where the systems waits for the nucleation of a cluster that triggers the emergence of polar order; $T_{\text{nuc}}$ and $T_{\text{stat}}$ are indicated by vertical dotted or dashed lines. \emph{Inset}:  for $\alpha=1$, no order develops, irrespective of the value of $\epsilon$ ($\epsilon=0, 5, 100$ from bottom to top).
{\bf (b)}  \emph{Free volume gain} $\Phi = \rho - \Gamma$ (triangles) for $\alpha = 5$ and $\epsilon=0.5$ exhibits a ``jump" at $\rho_c \approx 0.4$ (dashed line) and then is significantly above the reference curve $\rho- \Gamma_\text{hom}(\rho)$ (black $+$ are simulation results). $T_{\text{stat}}$ diverges for densities slightly above  $\rho_c$ (squares). 
{\bf (c)} \emph{Phase diagram} as a function of $\rho$ and $\epsilon$ for $\alpha=5$ (solid) and $\alpha=10$ (dashed).}
\end{figure*}

The pattern formation process can readily be quantified by introducing the time-dependent mean polarity ${\cal P} (t)$, i.e. the magnitude of the polarization averaged over all single particle polarizations $\vec{n}_i(t)$ contained in a region of interest 
%in the middle of the simulation box 
(ROI) or the entire simulation box (SIM), respectively. The polarization 
%$\vec{n}_i(t)$ 
of the $i$-th filament 
%at time $t$ 
is defined as the difference vector of the position of the filament's head between two successive time steps: $\vec{n}_i(t) := \vec{x}_{i}(t) - \vec{x}_{i}(t-1)$. These two measures of polarity provide complementary information about the pattern forming process, cf. Fig.~\ref{Fig2}d. ${\cal P}_\text{ROI}$ illustrates the formation of small polarized clusters that develop into bands: the sequence of spikes in the time traces indicate polar-moving clusters passing through the ROI 
%in the center of the simulation box. 
In contrast, ${\cal P}_\text{SIM}$ characterizes the evolution of global order in the system: the mean value slowly grows in time and fluctuations are much less pronounced. These latter fluctuations originate from changes in relative polarization of just a few large polar-moving structures with time. 
%(see also the videos in the Supplementary Material).

The temporal evolution of coherent motion can be accessed by quantifying the 
directional correlations in the filament's co-moving frame by means of the Lagrangian orientational correlation function:
\begin{equation}\label{pol_correlation}
	C(\tau; t) := \frac{ \sum_i \vec{n}_i(t) \cdot \vec{n}_i(t+\tau)  - \big( \sum_i \vec{n}_i(t)  \big)^2}{\sum_i \vec{n}_i^2(t)  - \big( \sum_i \vec{n}_i(t)  \big)^2} \, .
\end{equation}
The sum extends over all particles in the system, and $t$ denotes the waiting-time passed since the preparation of the system in a state with a uniform distribution in both, position and orientation. Starting from a disordered state at $t=0$, the directional correlation function decays exponentially in time $\tau$ with a decay length corresponding to the persistence length of the undisturbed persistent random walk $\ell_p^0$. As we increase the waiting-time $t$ a clear shoulder in $C(\tau;t)$ builds up until it finally becomes stationary; for the parameters used in Fig.~\ref{Fig2}e this occurs at $t \approx 5000$. Then, in the stationary regime, there are clearly two decay processes with well separated time scales: an initial fast decay and an extended time window with a significantly slower but again exponential decay. This indicates a phase separation where a fraction of the filaments is in a low density isotropic phase and the remainder is organized in coherently moving polar clusters. Filaments in the low density phase are responsible for the fast initial decay which has the same slope as the initial decay starting from an initially fully disordered filament configuration. In contrast, orientational correlations of filaments in clusters are significantly enhanced since they preferentially move into the same direction as  their neighbors due to the impact of the alignment field. This results in an increased decorrelation time of $C$. Since filaments are moving at unit velocity this time scale can also be interpreted as a persistence length of the trajectories traced out by the filaments, $\ell_p^\text{coll}$. We attribute the loss of correlations for filaments in clusters mainly to reorientations of the clusters as a whole and to a lesser extent to collisions between clusters. The latter and collisions of clusters with the confining boundary of the simulation box are responsible for the steep drop in the correlation function at very large times. The collective persistence length of the clusters $\ell_p^\text{coll}$ depend on the density $\rho$ as well as the strength of excluded volume interaction~$\epsilon$; see Supplementary Material. 
\begin{figure*}[t]
\centering
\includegraphics[width=18cm]{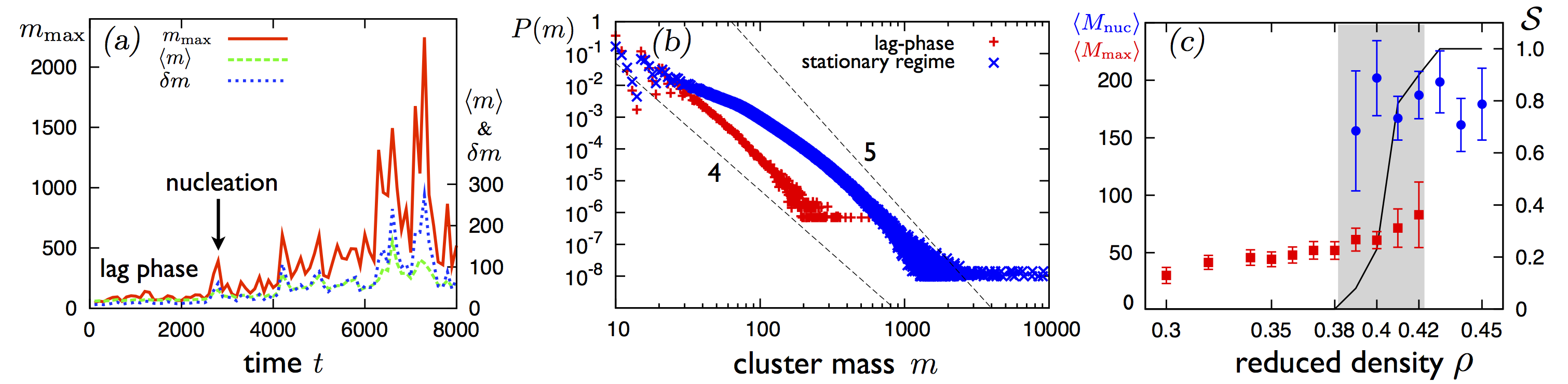}
\caption{\label{fig:cluster} (color online)
{\bf (a)} \emph{Cluster mass distribution} $P(m)$ for $\rho=0.6$ during the lag-phase and in the stationary regime of the polar ordered state sampled over more than $10^4$ realizations. For reference, power law distributions  $P(m)\propto m^{-\delta}$ with $\delta = 4 \, ,5$ are indicated by dashed lines.
{\bf (b)} \emph{Time traces} of $m_\text{max}$ [red solid], $\langle m \rangle$ [green dashed] and $\delta m$ [blue dotted] of the $P(m)$ for $\rho=0.6$. 
{\bf (c)} \emph{Mean values of the largest cluster}, $\langle M_\text{max} \rangle$ [red squares] and $\langle M_\text{nuc} \rangle$ [blue dots], serve as lower and upper bounds for critical nucleation cluster mass. In the transition window (shaded grey area) from the ISO to the PO phase, the fraction of realizations ${\cal S}$ which develop polar oder (solid line) increases from $0$ to $1$.  
Parameters for all graphs are $\alpha=5$ and $\epsilon=0.5$.}
\end{figure*}

To further analyze the self-assembly process and to characterize the ensuing stationary state we consider the fraction of occupied lattice sites $\Gamma$, or equivalently the free volume fraction $\Phi := \rho - \Gamma$. In the absence of polar alignment processes between the filaments, we find that $\Gamma$ fluctuates around a constant value depending on the strength of the excluded volume interaction $\epsilon$ (see Fig.~\ref{fig:transition}a inset).  For a non-interacting system ($\epsilon=0, \alpha=1$), the stationary value of $\Gamma$ as a function of reduced density $\rho$ is given by $ \Gamma_\text{hom} (\rho) =1 - e^{-\rho}$; see Fig.~\ref{fig:transition}b. This result has previously been obtained in continuum percolation theory for the fraction of lattice sites occupied by statistically homogeneous distribution of overlapping rods~\cite{Torquato}. In the presence of excluded volume interaction, the actual value of $\Gamma$ is only slightly smaller than $\Gamma_\text{hom}$ and shows the same functional form. Adding polar alignment interactions leads to qualitatively different behavior. As can be inferred from Fig.~\ref{fig:transition}b, there is now a threshold density $\rho_c$, where the free volume gain $\Phi$  jumps to a value much larger than the corresponding value for a non-interacting system. Moreover, measuring the polar order parameters it turns out that this jump actually coincides with the onset of polar order, cf. Fig.~\ref{Fig2}e. The jump in $\Phi$ may, therefore, be taken as a signature to map out the phase diagram as shown in Fig.~\ref{fig:transition}c; for a more detailed discussion see the Supplementary Material. It also clearly indicates that filaments must have formed some clusters much denser than expected for a purely statistical overlap of filaments. Figure \ref{fig:transition}a shows that clusters form by a nucleation process: There is a lag-phase during which $\Gamma$ stays high close to a value obtained in the absence of polar alignment processes. Subsequently, there is a time window $[T_\text{nuc}, T_\text{stat}]$, where the available free volume fraction $\rho - \Gamma$ increases towards a higher stationary value, cf. Fig.~\ref{fig:transition}b. The time to reach the final polar steady state  diverges as a power law $T_\text{stat} \propto (\rho - \rho_c)^{-\zeta}$ with $\zeta \approx 1$.

% Discussion of the phase diagram:
%\note{The phase behavior of the model comprises two states, an isotropic (ISO) state and a polar-ordered (PO) state, cf. Fig.~\ref{fig:transition}c. As expected, for a fixed value of the interaction parameters, there is a critical density $\rho_c$ above which polar order is obtained. Interestingly, for a given density, the polar ordered state breaks down if the excluded volume interaction exceeds a critical value. This value becomes smaller with decreasing alignment interaction. Actually, for rods interacting by excluded volume only ($\alpha = 1$) we find no polar-ordered state,  consistent with previous findings for off-lattice models of overdamped hard rods \cite{Peruani_rods}. We conclude that alignment interaction ($\alpha > 1$) is a necessary precondition for the existence of a phase transition to a macroscopically ordered polar state. There is a critical threshold value for $\alpha_c (\rho, \epsilon)$ which remains larger than $1$ even in the absence of excluded volume interaction; see Supplementary Material. In the absence of polar alignment interaction, our model shows only an isotropic phase but no nematic phase as recently reported in Ref.~\cite{Ginelli} studying a Viscek-like model with a nematic alignment rule. This may be attributed to a lack of excluded volume interaction in Ref.~\cite{Ginelli, Note1}.}

To further scrutinize the nucleation processes, we analyzed the cluster mass distribution $P(m)$ and the size of the critical nucleation cluster as a function of density,  cf. Fig.~\ref{fig:cluster}b and c. Here clusters are defined as connected areas on the lattice with local occupation numbers $k \ge k_c = 5$, see Supplementary Material. For an illustration, Figure~\ref{fig:cluster}a shows time traces for the maximum of the cluster mass distribution $m_{\text{max}}$, the mean $\langle m \rangle$ and standard deviation $\delta m = \sqrt{ \langle m^2 \rangle -\langle m \rangle^2}$ close to the phase transition from an isotropic to an ordered state. We observe that during the lag-phase clusters are continually created and destroyed such that all cluster characteristics remain constant on average and there is no gain in available free volume. The corresponding distribution of cluster sizes $P(m)$ shows power law tails with an exponent close to $\delta=5$, cf. Figure~\ref{fig:cluster}b.
For the representative trace shown in Fig.~\ref{fig:cluster}a, there is a spontaneous nucleation of a large cluster at $t \approx 2800$, that albeit it partially disassembles still seems to trigger the transition to cluster formation. It defines the onset of a steady increase in $\langle m \rangle$ and also a change in the power law of the cluster size distribution, cf.  Figure~\ref{fig:cluster}b. In conclusion, close to the critical point the system is waiting for a large enough density fluctuation that acts as a nucleus for cluster formation, which in turn triggers the transition to the polar ordered state; as shown above the waiting time diverges upon approaching the transition, cf. Fig.~\ref{fig:transition}b.

To quantify the size of the critical nucleation cluster and further substantiate  that the nucleation phenomenon is the mechanism driving the transition we have performed a detailed analysis of the largest cluster formed during the nucleation process and thereby quantified the heuristic observations made in Fig.~\ref{fig:cluster}a. In the isotropic phase, one expects that the largest clusters formed over the whole observation window $M_\text{max} :=  \underset{\text{t}}{\text{max}} \; m_\text{max} (t)$ are typically too small to act as a nucleus for the emergence of polar order. However, in a finite system, there will still be rare occasions where fluctuations are strong enough to form a critical nucleus. The fraction ${\cal S}$ of realizations which develop polar order sharply increases as the density is raised throughout the transition window (see Fig. \ref{fig:cluster}c), with the critical density $\rho_c \approx 0.4$.  Below the critical density, the values of $M_\text{max}$ for those realizations which did not lead to polar order are a measure of cluster sizes which are too small to trigger the pattern forming process, and can thereby be used as lower bounds for the size of the critical nucleus. Figure~\ref{fig:cluster}c shows the average value of $M_\text{max}$ for those realizations which did not lead to polar order.  We observe that this quantity increases upon approaching the critical density from below.  Above the critical density, most realizations will develop polar order and one may estimate the critical nucleus size as the mean size of the largest cluster at the nucleation time $ M_\text{nuc} := m_\text{max} (t=T_\text{nuc})$. The corresponding average value, $\langle M_\text{nuc} \rangle$ gives the typical cluster mass that has been large enough to trigger a transition to polar order, and hence serves as an upper bound for the critical nucleus.

While previous numerical studies on propelled particle systems focussed on the long-time dynamics, we investigated the dynamics of the self-assembly process leading to the emergence of polar ordered states. We find that the ordering process is driven by the formation of a critical nucleation cluster: once a cluster of sufficiently large mass has assembled, the entire system builds up collective motion. This is reflected in the existence of a lag time, where the system does not succeed to assemble a cluster of sufficient mass. The signatures of the ensuing phase transitions are a jump in the free volume and a power-law divergence in the characteristic time for the formation of the polar ordered state.  We suggest that exploring the principles underlying the self-assembly process of polar clusters is a promising route to understand ordering phenomena in a broader class of active systems.

\begin{acknowledgments} The authors would like to thank Fernando Peruani for helpful discussions. Financial support from the DFG in the framework of the SFB 863 and the German Excellence Initiative via the program Nanosystems Initiative Munich (NIM) is gratefully acknowledged.
\end{acknowledgments}

%
%\bibliographystyle{prsty}
%%\bibliographystyle{unsrt}
%%\bibliographystyle{prsty}
%
%%\bibliography{Nucleation-induced transition to collective motion in active systemsNotes}
%\bibliography{References}
%
%

%%%%%%%%

%\documentclass[aps,prl,preprint,superscriptaddress]{revtex4}

\end{document}